\documentclass[aps,prl,reprint,superscriptaddress,footinbib,floatfix]{revtex4-2}

\usepackage{graphicx}
\usepackage{bm}
\usepackage{amssymb}
\usepackage{amsbsy}
\usepackage{amsmath}
\usepackage{mathtools}
\usepackage{xcolor}
\usepackage{stmaryrd}
\usepackage{mathrsfs}
\usepackage[mathscr]{euscript}
\usepackage{tikz}
\usepackage{float}
\usetikzlibrary{shapes}
\usepackage{cancel}
\usepackage[normalem]{ulem}

\begin{document}


\title{Migration of Active Rings in Porous Media}


\author{Ligesh Theeyancheri}
\affiliation{Department of Chemistry, Indian Institute of Technology Bombay, Mumbai, Maharashtra -  400076, India}
\author{Subhasish Chaki}
\affiliation{Department of Chemistry, Indian Institute of Technology Bombay, Mumbai, Maharashtra -  400076, India}
\author{Tapomoy Bhattacharjee}
\email{tapa@ncbs.res.in}
\affiliation{National Centre for Biological Sciences, Tata Institute of Fundamental Research, Bangalore}
\author{Rajarshi Chakrabarti}
\email{rajarshi@chem.iitb.ac.in}
\affiliation{Department of Chemistry, Indian Institute of Technology Bombay, Mumbai, Maharashtra -  400076, India}



\begin{abstract}
\noindent Inspired by how the shape deformations in active organisms help them to migrate through disordered porous environments, we simulate active ring polymers in two-dimensional random porous media. Flexible and inextensible active ring polymers navigate smoothly through the disordered media. In contrast, semiflexible rings undergo transient trapping inside the pore space; the degree of trapping is inversely correlated with the increase in activity. We discover that flexible rings swell while inextensible and semiflexible rings monotonically shrink upon increasing the activity. Together, our findings identify the optimal migration of active ring polymers through porous media.
\end{abstract}


\maketitle


\section{Introduction}

\noindent Active agents live and move through the disordered porous environment. The pores serve as selective-permeability barriers in regulating diffusive transport that play a crucial role in tissue protection and cell functioning of human and animal bodies~\cite{cherstvy2019non,witten2017particle,kumar2019transport,bucci2020systematic,shin2015polymer,moretta2002natural,daher2018next,ferlazzo2012natural,parkhurst1994leukocytes}. At the cellular level, when a cell touches a permissive surface, it will form adhesive structures to cope up with the hostile environments~\cite{de2017single,alert2020physical,buttenschon2020bridging,mitrossilis2009single,moreira2020adhesion}. This type of adaptive migration has been observed in different types of cells; amoeba~\cite{pollard1970cytoplasmic, hu2017liquid}, neutrophils, bacterial cells~\cite{bhattacharjee2021}, and dendritic cells~\cite{friedl2008interstitial,renkawitz2019nuclear}. In bio-remediation, microscopic organisms migrate through the porous media such as soil and sediments to remove or neutralize the environmental pollutants by metabolic processes~\cite{adadevoh2016chemotaxis,adadevoh2018chemotaxis,ford2007role}. The physical properties of the microenvironment play a significant role in the migration of these living objects across all sizes starting from single cells--- substrate stiffness and pore size regulates the migration of eukaryotic cells \cite{pathak2012independent,charras2014physical,harley2008microarchitecture, zaman2006migration,vernerey2021mechanics,braig2015pharmacological}--- to multicellular organisms---nematodes swim in low viscosity fluids whereas they navigate by crawling on soft surfaces \cite{fang2010biomechanical,shen2012undulatory}. To migrate effectively through crowded disordered environments, cells deform their shapes\cite{bhattacharjee20183d,lammermann2008rapid,renkawitz2009adaptive,aoun2021leukocyte,hu2019high} by consume energy from ATP hydrolysis in two ways; by the action of myosin molecular motors on actin filaments~\cite{friedl2003tumour, rorth2009collective, sens2020stick} or by a propulsive mechanism using the plasma membrane blebs~\cite{fackler2008cell}. Hence investigating the mechanism behind the migration of these active agents through the complex porous environments will help us to understand the fundamental physiological and pathological processes. \\ 

\noindent In unconfined liquid media, a single active particle exhibits transient super diffusion followed by a long-time enhanced diffusion~\cite{bechinger2016active,wu2000particle}. However, novel non-equilibrium effects emerge in the conformational and dynamical properties of a chain of interlinked active particles. For flexible active polymers, the polymer chain swells with increasing the activity~\cite{chaki2019enhanced,samanta2016chain,osmanovic2017dynamics,shin2015facilitation}. In contrast, semiflexible active polymers shrink at low activity and swell for large activity~\cite{eisenstecken2016conformational}. However, much less is explored about how the topology of the porous media affects the migration of active agents~\cite{wu2021mechanisms}. Studies of linear active polymers in two-dimensional periodic porous medium demonstrated that the stiff chains are able to move almost unhindered through the ordered porous medium whereas the flexible one gets stuck~\cite{mokhtari2019dynamics}. In biological systems, cell morphology can have a crucial impact on different modes of migration in porous environments. Recent studies have shown that the micro-confinement of the porous medium dramatically alters the run and tumble motion of rod shaped bacterial cells to hopping and trapping motion~\cite{bhattacharjee2019bacterial,kurzthaler2021geometric}. On the other hand, the amoeboid cells, which are typically roundish, are known to migrate fast and adapt quickly to their surroundings~\cite{bhattacharjee20183d}. Thus, active ring polymer can also be a model for mammalian cells~\cite{gupta2021role,teixeira2021single}. Dynamical and conformational properties of active ring polymers in complex environments remain largely elusive and could be markedly different from active linear polymer counterparts despite of having an equal number of active monomers in both cases~\cite{chubak2022active,locatelli2021activity}. But there are no systematic study of such an active ring polymer in random porous environments. \\ 

\noindent In this work, we explore the dynamics of active ring polymers in two-dimensional random porous media. We extensively analyze the dynamical and conformational properties of the active ring polymer by following the position ($r_c$) of the center of mass (COM) and the radius of gyration ($\text{R}_g$). In general, we find that the dynamics of the COM of the ring polymer in porous environment is always enhanced due to the combined effects of activity and conformational fluctuations of the polymer. Interestingly, the semiflexible ring polymer shows a transition from trapped state to escaping with increasing the activity at longer times. Whereas such trapping events are absent in the case of flexible ring polymers. Moreover, upon increasing the activity, the distribution of $\text{R}_g$ begins to broaden towards larger values of $\text{R}_g$ indicating that the activity drives the flexible ring polymers to continuously swell. However, for the inextensible and semiflexible ring polymers, the distribution of $\text{R}_g$ broadens and shifts towards lower values of $\text{R}_g$ as a function of activity. This is indicative of shrinking of the active ring polymer. As a comparative study, the pore size effect is also specifically investigated. Hence, studies of active ring polymers in porous environments will help in designing the active migration strategies of shape deforming systems.

\section{Method}

\subsection{Model and Simulation Details}

\begin{figure*}
	\centering
		\includegraphics[width=0.7\linewidth]{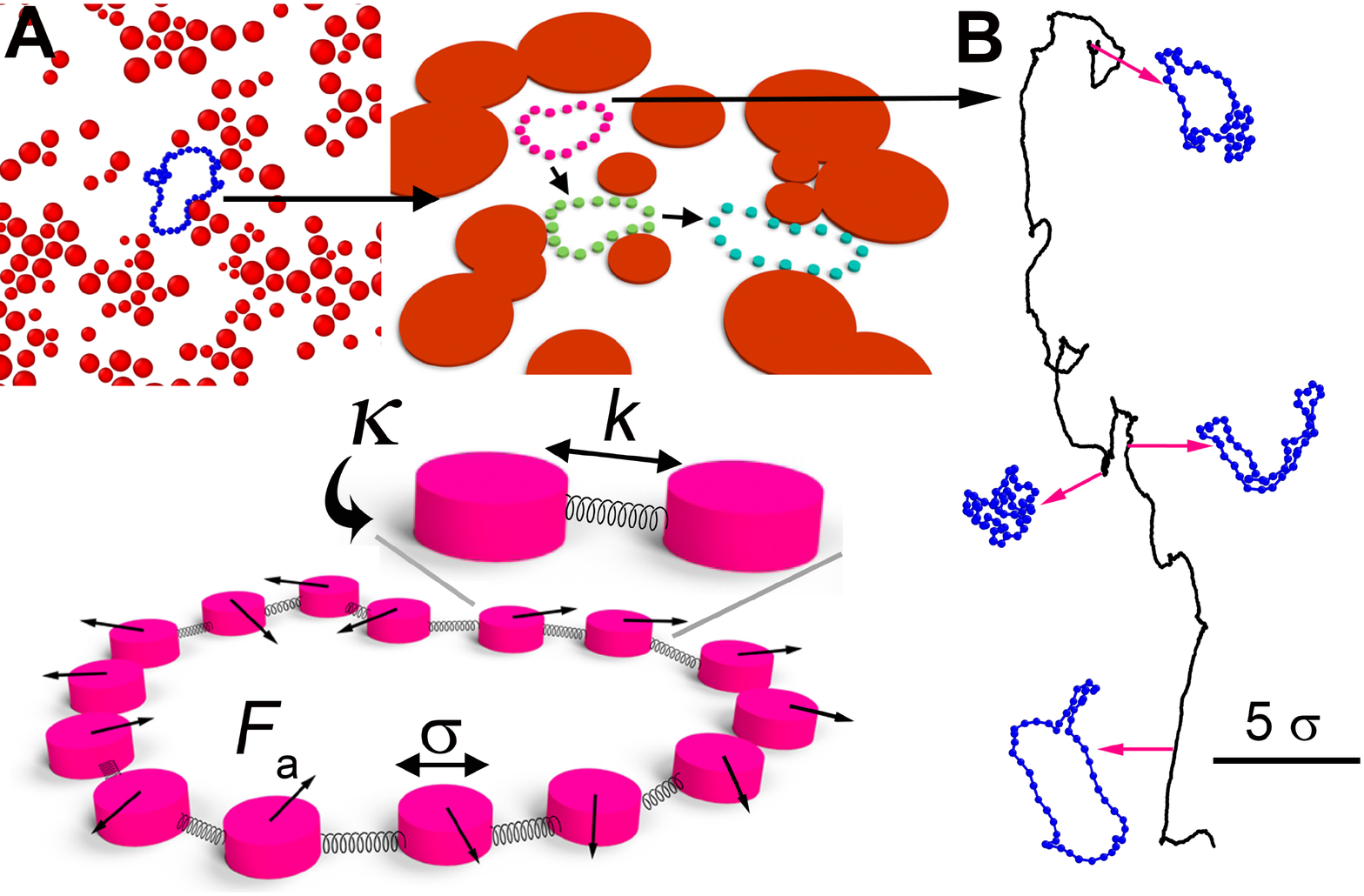} \\
\caption{(A) A snapshot of active ring polymer in porous media and the schematic sketch of active ring polymer. Polymer is regarded as beads connected by springs. (B) A representative trajectory of the active ring polymer in the porous medium. The magenta arrows represent ring conformations corresponding to different frames in the trajectory.}\label{fig:model_traj}
\end{figure*}

\noindent We model the polymer ring as a sequence of N self-propelled beads of diameter $\sigma$ connected by N finitely extensible springs (Fig.~\ref{fig:model_traj}A). The dynamics of the ring polymer is determined by the evolution of the self-propelled beads of positions $r_{i}$ which we simulate using Langevin dynamics, 
\begin{equation}
m\frac{d^2 \textbf{r}_{i}(t)}{dt^2} = - \gamma \frac{d \textbf{r}_{i}}{dt} - \sum_{j} \bigtriangledown V(\textbf{r}_i-\textbf{r}_j) + {\bf f}_{i}(t) + {\bf{F}_{\text{a, i}}(t)}
	\label{eq:langevineq}
\end{equation} 
where the drag force, $\gamma \frac{d \textbf{r}_{i}}{dt}$ is the velocity of each bead times the friction coefficient $\gamma$, $m$ is the mass of monomer, $V(r)$ is the total interaction potential which gives rise to the conservative forces, thermal force $\bf f_{i}(t)$ is modeled as Gaussian white noise with zero mean and variance $\left<f_{i}(t^{\prime})f_{j}(t^{\prime\prime})\right> = 4 \gamma k_B T \delta_{ij}\delta(t^{\prime}-t^{\prime\prime})$, and ${{F}_{\text{a, i}}(t)}$ is the active force which drives the system out of equilibrium. ${\bf{F}_{\text{a, i}}(t)}$ has the magnitude $F_{\text{a}}$, acts along the unit vector~\cite{bechinger2016active} of each $i^{th}$ monomer, $\bf{n}(\boldsymbol{\theta_i}) = {(\textrm{cos} \,\boldsymbol{\theta_i}, \, \textrm{sin} \, \boldsymbol{\theta_i})}$ , where $\theta_i$ evolves as $\frac{d \boldsymbol{\theta_i}}{dt} = \sqrt{2D_R} {\bf f}_{i}^{R}$, $D_R$ is the rotational diffusion coefficient and ${\bf f}_{i}^{R}$ is the Gaussian random number with a zero mean and unit variance. Hence, the persistence time of the individual monomers is related to the rotational diffusion coefficient, $D_R$ as $\tau_R = \frac{1}{D_R}$. We measure the distance in the unit of diameter of the monomers of the ring polymer $\sigma$, energy in $k_{B}T$, and time in $\tau=\sqrt{\frac{m \sigma^2}{k_{B}T}}$. The total interaction potential $V(r) = V_{\text{FENE}} + V_{\text{BEND}} + V_{\text{WCA}}$ consists of bond, bending and excluded volume contributions. The bond stretching is controlled by a FENE potential:
 \begin{equation}
V_{\text{FENE}}\left(r_{ij}\right)=\begin{cases} -\frac{k r_{\text{max}}^2}{2} \ln\left[1-\left( {\frac{r_{ij}}{r_{\text{max}}}}\right) ^2 \right],\hspace{3mm} \mbox{if } r_{ij} \leq r_{\text{max}}\\
\infty, \hspace{35mm} \mbox{otherwise}.
\end{cases}
\label{eq:FENE}
\end{equation}
where $r_{ij}$ is the distance between two neighboring monomers in the ring polymer with a maximum extension of $r_{\text{max}} = 1.5 \sigma$, and $k$ is the spring constant~\cite{kremer1990dynamics}. To achieve the condition of inextensibility,  $k$ is set to be very high for the inextensible ring polymer ($k = 1000$). The stiffness of the ring polymer is implemented through the bending potential,
 \begin{equation}
V_{\text{BEND}}\left(\phi_i \right) = \kappa \left(1-\cos\phi_i\right)
\label{eq:BEND}
\end{equation}
where $\kappa$ is the bending modulus and $\phi_{i}$ is the angle between the bond vectors $i$ and $i+1$. To account for self-avoidance a pair of monomers of the ring polymer interact $via$ the repulsive Weeks–Chandler–Andersen (WCA) potential~\cite{weeks1971role}.
\begin{equation}
V_{\textrm{WCA}}(r_{ij})=\begin{cases}4\epsilon_{ij} \left[\left(\frac{\sigma_{ij}}{r_{ij}}\right)^{12}-\left(\frac{\sigma_{ij}}{r_{ij}}\right)^{6}\right]+\epsilon, \mbox{if }r_{ij}<2^{1/6}\sigma_{ij} \\
0, \hspace{35mm} \mbox{otherwise},
\end{cases}
\label{eq:WCA}
\end{equation}
where $r_{ij}$ is the separation between the interacting particles, $\epsilon_{ij} = 1$ is the strength of the steric repulsion, and $\sigma_{ij} = \frac{\sigma_i + \sigma_j}{2}$ determines the effective interaction diameter, with $\sigma_{i(j)}$ being the diameter of the interacting pairs. The porous medium is modeled by randomly placing $M$ $(M=1200, 2000, 2500, 3000)$ number of particles that are allowed to overlap inside a two-dimensional square box of length 300 $\sigma$. The size of the beads forming the porous medium, $\sigma_P$ ranges from 1 to 10 $\sigma$, and the size distribution of these particles fall under a Gaussian distribution of mean $6$ $\sigma$. These beads are static throughout the simulations and interact with the monomers of ring polymer through WCA potential (Eq.~\ref{eq:WCA}). We have consider three different cases: flexible ($k = 30$ and $\kappa = 0$), inextensible  ($k = 1000$ and $\kappa= 0$), and semiflexible ($k = 30$ and  $\kappa = 1000$) ring polymers in the porous media.\\

\noindent All the simulations are performed using the Langevin thermostat and the equation of motion is integrated using the velocity Verlet algorithm in each time step. We initialize the system by randomly placing the ring polymers inside the porous medium and relaxed the initial configuration for $2 \times 10^6$ steps. All the production simulations are carried out for $10^9$ steps where the integration time step is considered to be $10^{-5}$ and the positions of the monomers are recorded every $100$ steps. The simulations are carried out using LAMMPS~\cite{plimpton1995fast}, a freely available open-source molecular dynamics package.
  
\subsection{Characterization of the porous media}

\begin{figure}
\centering
   \includegraphics[width=0.99\linewidth]{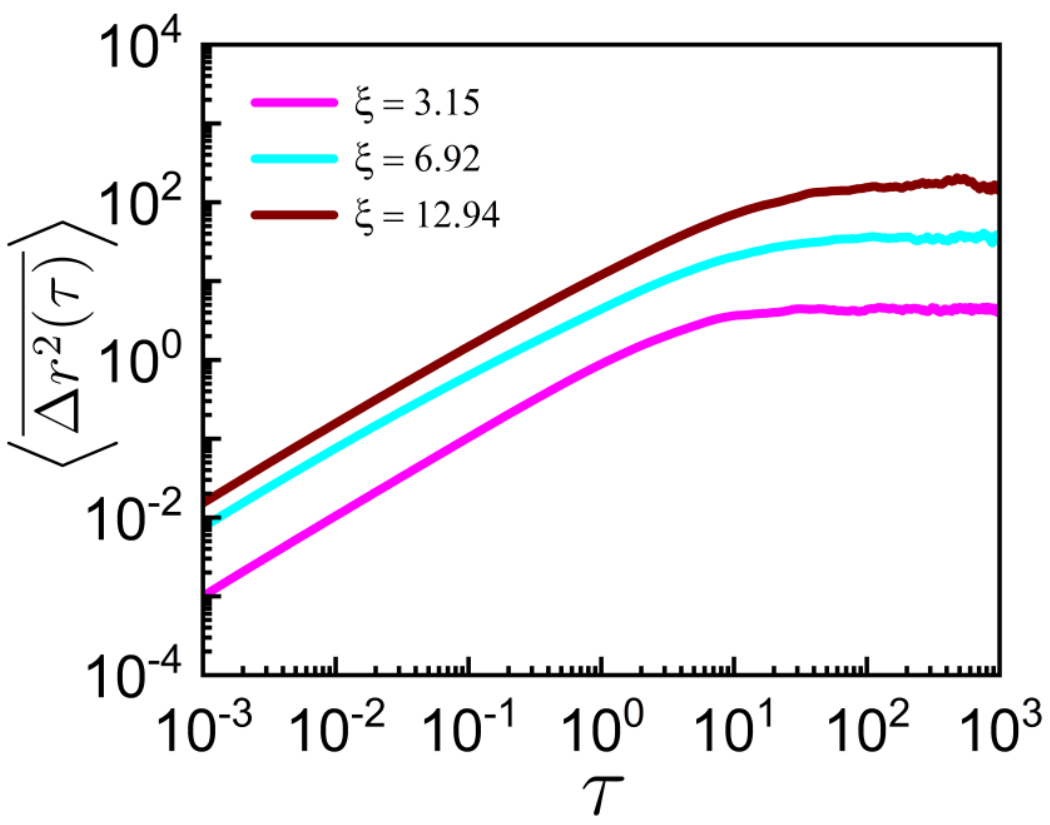}\\
\caption{Log-log plot of $\left\langle{\overline{\Delta r^{2}(\tau)}}\right\rangle$ $vs$ $\tau$ of the passive tracer particle in different porous media.}\label{fig:probe_MSD}   
\end{figure}

\noindent We characterize the pore space structure by the average pore size $\xi$. We place passive tracer particles at different random locations in the media and allow them to diffuse through the polydisperse porous media. The average pore size $\xi$ is calculated from the time-and-ensemble averaged mean square displacements (MSD) of the passive tracer particle in the porous medium (Fig.~\ref{fig:probe_MSD}). At a longer time, the tracer is confined by the obstacles, and the MSD saturates. Then we take the square root of this saturated value and add the tracer particle diameter to get $\xi$. The values of $\xi$ obtained from different tracer trajectories are binned to construct a histogram from which the ensemble-averaged probability distribution of $\xi$, $\text{P}(\xi)$ is computed (Fig.~S1). To change the average pore size, $\xi$, we increase the obstacle density by adding particles, keeping the same width of the Gaussian distribution of the sizes of the obstacles. The pore sizes of these different porous media are less/bigger than or comparable to the average $\text{R}_g$ of the passive ring polymer in unconfined space. 

\section{Results}

\noindent We first investigate the dynamics of flexible ring polymers ($\kappa = 0$) with a small spring constant of $k = 30$ to find that active motion drives instantaneous deformations in the polymer rings; tracking the center of mass ($r_c$) of each rings reveals that the optimal motility of these rings is facilitated by a series of conformational changes (Fig.~\ref{fig:model_traj}B and Movie S1). To further quantify the dynamical behavior of the active ring polymer, we analyze the time-and-ensemble averaged Mean Squared Displacement (MSD), $\left\langle{\overline{\Delta r_{c}^{2}(\tau)}}\right\rangle$, of the center of mass as a function of lag time $\tau$. First we compute the time-averaged MSD, $\overline{\Delta r_i^{2}(\tau)} = \frac{1}{T^{\prime}-\tau} \int_{0}^{T^{\prime}-\tau} {\left[ \textbf{r}_i(t+\tau) - \textbf{r}_i(t)\right]}^2  dt$ from the time series $\textbf{r}_i(t)$. Here, $T^{\prime}$ is the total run time and $\tau$ is the lag time (width of the window slide along a single trajectory for averaging). To obtain the time-and-ensemble-averaged MSD, we compute the average, $\left\langle{\overline{\Delta r^{2}(\tau)}}\right\rangle  =  \frac{1}{N^{\prime}} \sum_{i=1}^{N^{\prime}}{\overline{\Delta r_i^{2}(\tau)}}$, where $N^{\prime}$ is the number of independent trajectories. If measured time series are not long enough, $\left\langle{\overline{\Delta r^{2}(\tau)}}\right\rangle$ provides smoother curve when $N^{\prime}$ is sufficiently large. For a given set of  parameters we generate $20$ independent trajectories of the ring polymer. Typically, $\left < {\overline{\Delta r^{2}(\tau)}}\right >$ scales with $\tau$ as $\sim \tau^{\alpha}$, where  $\alpha(\tau)=\frac{d \log \left<\overline{\Delta r^{2}(\tau)}\right>}{d \log \tau}$. The scaling exponent $\alpha$ determines the type of diffusion: $\alpha = 1$ corresponds to normal diffusion, $\alpha < 1$ corresponds to sub-diffusion, and $\alpha > 1$ corresponds to super-diffusion. In the absence of any active force and in unconfined environment, MSD varies linearly with time reflecting the over-damped dynamics of the ring polymer (Fig.~S2). In unconfined space, the MSD of the ring polymers exhibit three distinct regimes of motion in the presence of activity: a short time thermal diffusion, intermediate super-diffusion, and an enhanced diffusion at longer times as compared to the passive polymers~\cite{wu2014three,eisenstecken2017internal,bhattacharjee20183d,ghosh2014dynamics,chaki2019enhanced,eisenstecken2016conformational,osmanovic2017dynamics}. The intermediate superdiffusion occurs at an earlier time with an increase in the activity. A similar trend in the translational dynamics of the center of mass is observed in analytical studies of ideal free active linear polymer~\cite{chaki2019enhanced,eisenstecken2017internal} and ring polymer ~\cite{mousavi2019active} where the activity is modeled as an Ornstein-Uhlenbeck process.

\subsection{Migration of flexible, inextensible and semiflexible rings in the porous media: smooth migration $vs$ trapping}

\begin{figure*}
	\centering
		\includegraphics[width=0.95\linewidth]{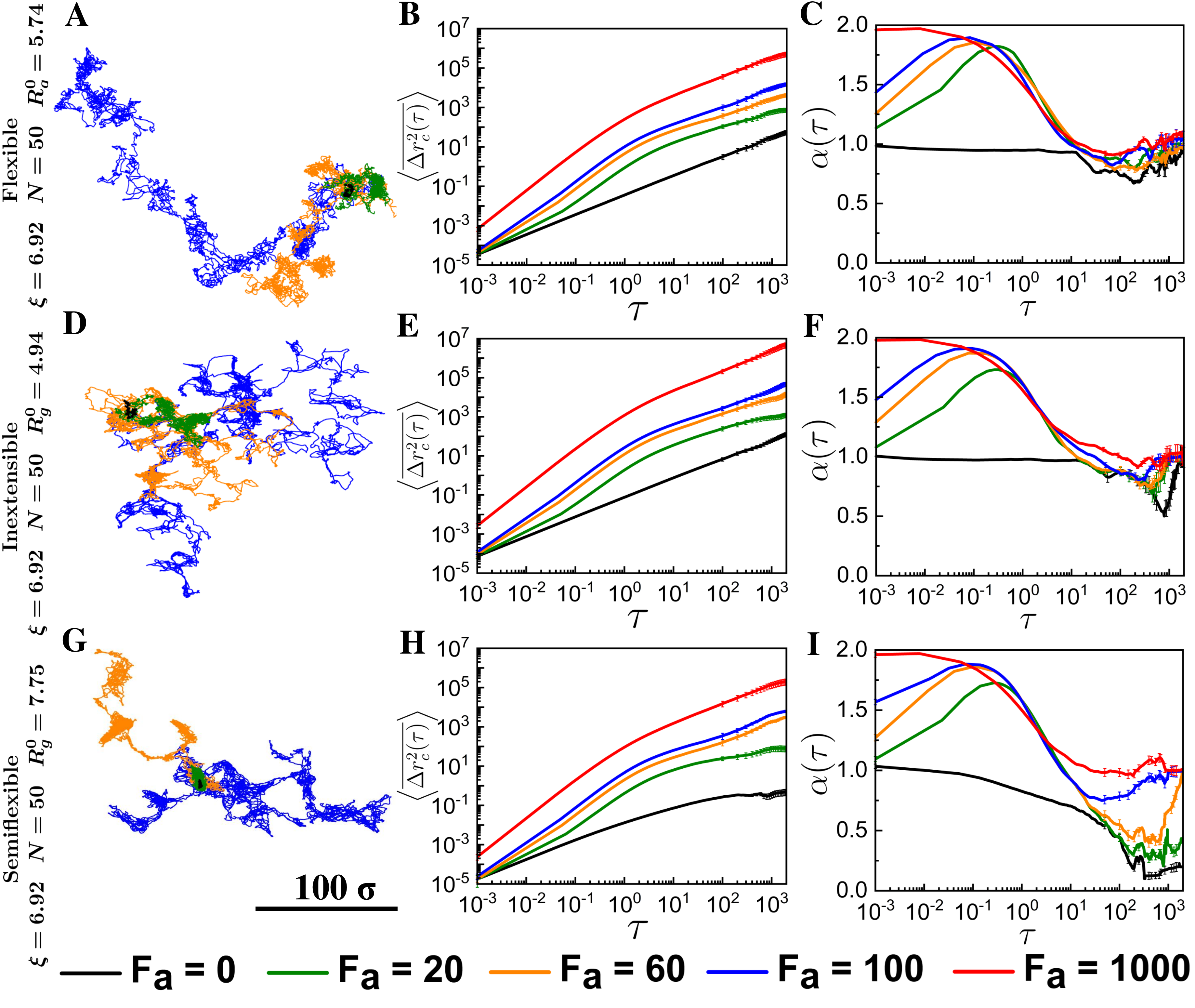} 
\caption{The COM trajectory, log-log plot of $\left\langle{\overline{\Delta r_{c}^{2}(\tau)}}\right\rangle$ $vs$ $\tau$, and log-linear plot of $\alpha(\tau)$ $vs$ $\tau$ of the ring subjected to different $\text{F}_a$ for flexible (A, B, and C), inextensible (D, E, and F), and semiflexible (G, H, and I) ring polymers respectively in the porous medium with $\xi = 6.92$. For each time interval $\tau$, the error bars corresponding to standard deviation are plotted, even if too small to be visible. N is the number of monomers of the ring polymer and $\text{R}_g^0$ is the average radius of gyration of the passive ring polymer in unconfined space.}\label{fig:dynamics_msd}
\end{figure*}

\noindent To further investigate the dynamics of the rings under confinement, we simulate motions of an active ring polymer inside disordered poly-disperse porous media modeled by randomly placing $M$ number of circular obstacles that are allowed to overlap inside a two-dimensional square box of fixed size. The size of the obstacles forming the porous media follow Gaussian distribution. These obstacles are static throughout the simulations and interact with the monomers of ring polymer $via$ the repulsive Weeks–Chandler–Andersen (WCA) potential~\cite{weeks1971role}. Different porous media with varying void volumes are prepared by varying the number of obstacles $M$ $(M=1200, 2000, 2500, 3000)$. We characterize the extent of confinement offered by these porous media by the average pore size, $\xi$, measured for each porous medium by identifying the average cage size for a thermal fluctuating tracer (Fig.~\ref{fig:probe_MSD}). Together, these measurements indicate the variation in the extent of confinement. We carry out independent simulations by randomly placing the ring polymer inside the pore space. \\ 

\begin{figure}  
\centering
   \includegraphics[width=0.9\linewidth]{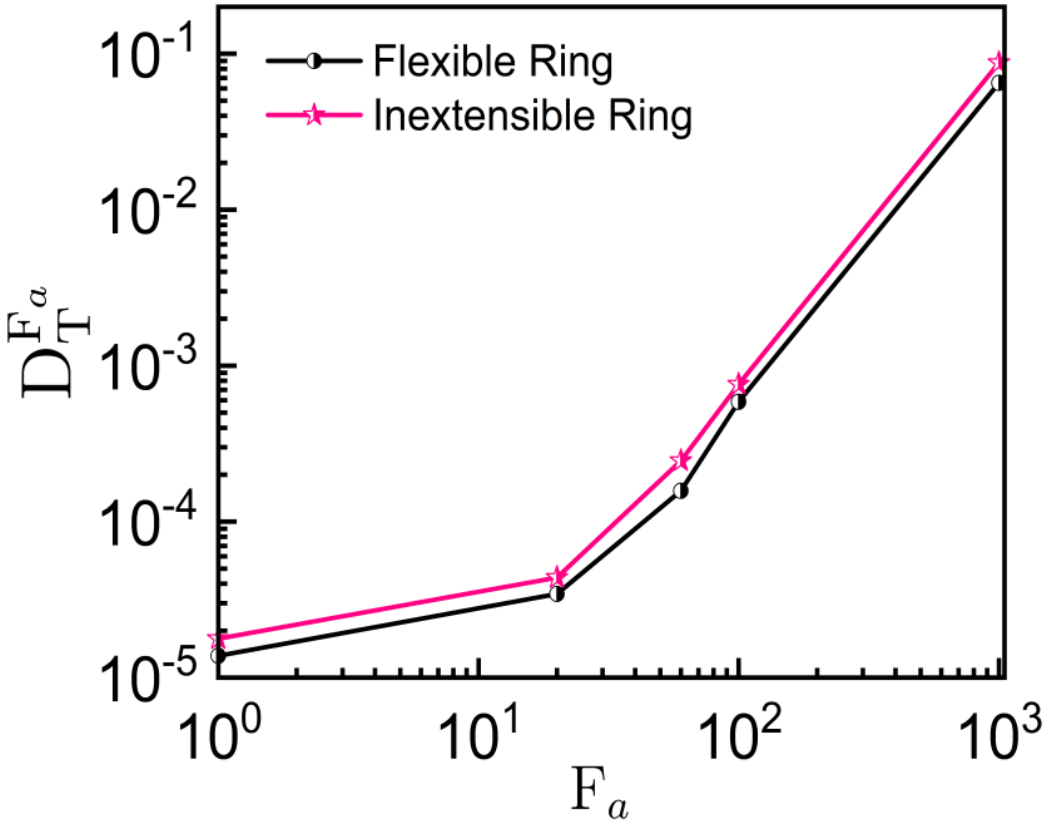}\\
\caption{The effective translational diffusion coefficient ($\text{D}_\text{T}^{\text{F}_a}$) of the flexible and inextensible ring polymers (N = 50) subjected to different activity in porous medium with $\xi = 6.92$.}\label{fig:diffusivity}
\end{figure}

\noindent In disordered environment, the dynamics of the flexible ring polymers is altered and is set by the overall activity and the extent of confinement (Fig.~\ref{fig:dynamics_msd}A and Movie S2). Primarily, a transient behavior between two regimes of motion emerges: a short-time super-diffusion which occurs due to active motion, and a long-time diffusive behavior (Fig.~\ref{fig:dynamics_msd}(B, C)). However, in case of inextensible rings ($k = 1000$, $\kappa = 0$) an intermediate subdiffusive behavior emerges for passive and lower activities followed by the diffusive behavior at long-time (Fig.~\ref{fig:dynamics_msd}(D, E, and F)). The rings get caged intermittently (Movie S3). We extract the values of effective translational diffusion coefficients ($\text{D}_\text{T}^{\text{F}_a}$) of the flexible and inextensible active ring polymers by fitting the long time $\left\langle{\overline{\Delta r_{c}^{2}(\tau)}}\right\rangle$ with $4\text{D}_\text{T}^{\text{F}_a}\tau$. In Fig.~\ref{fig:diffusivity}, we plot $\text{D}_\text{T}^{\text{F}_a}$ as a function of $\text{F}_a$. The diffusivity of the rings increase with increasing activity in the porous medium and for the inextensible ring polymer has a larger diffusivity as compared to the flexible ring for a fixed $\text{F}_a$. This implies that the activity always help the ring polymers to navigate more efficiently through the porous media.
Finally, we investigate the dynamics of semiflexible ring polymers ($\kappa = 1000$) with a small spring constant of $k = 30$. The high bending rigidity restricts the conformational fluctuations leading to the trapping of these ring polymers in tight spaces of the porous environment (Fig.~\ref{fig:dynamics_msd}G). For small activities, long-time sub-diffusive behavior is more pronounced in the dynamics of the semiflexible rings (Fig.~\ref{fig:dynamics_msd}(H, I)). However, an increase in activity enhances the conformational fluctuations of the semiflexible ring polymers, which helps them to escape from the micro-confinements of the porous medium (Movie S4). This behavior is illustrated in Fig.~\ref{fig:dynamics_msd}G where for constant $\text{F}_a$, the path traversed by the COM of semiflexible ring polymer is reduced compared to that of the flexible and inextensible rings. We find that the dynamical behavior of the flexible ring polymer as observed in Fig.~\ref{fig:dynamics_msd}A persists for a broad range of $\xi$, but becomes rather enhanced for higher $\xi$ as exemplified by the different trajectories in Fig.~S3(A and B). At the short and intermediate time, $\alpha(\tau)$ remains independent of the porous structure of the environment (Fig.~S3C). At longer time, $\alpha(\tau)$ exhibits an increase with $\xi$ indicates the polymer exploring different random pores in the medium.  

\subsection{Conformations of rings in unconfined and porous media: activity induced swelling and shrinking}

\begin{figure*}
	\centering
		\includegraphics[width=0.95\linewidth]{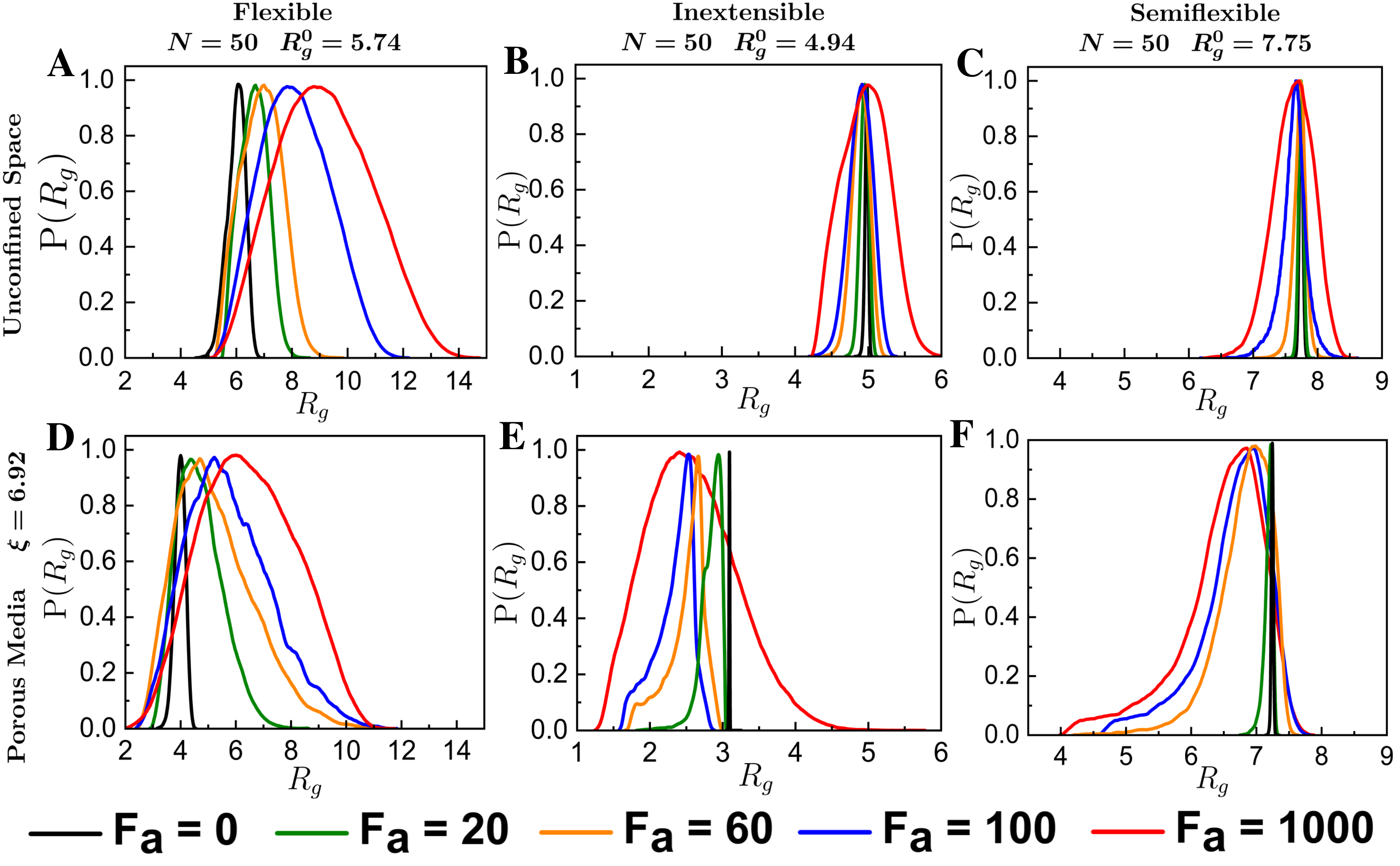} 
\caption{$\textrm{P}(\text{R}_g)$ $vs$ $\text{R}_g$ for flexible (A and D), inextensible (B and E), and semiflexible (C and F) ring polymers ($N = 50$) subjected to different activity in the unconfined space and porous medium with $\xi = 6.92$ respectively.}\label{fig:conformation}
\end{figure*}

\begin{figure*}
	\centering
		\includegraphics[width=0.95\linewidth]{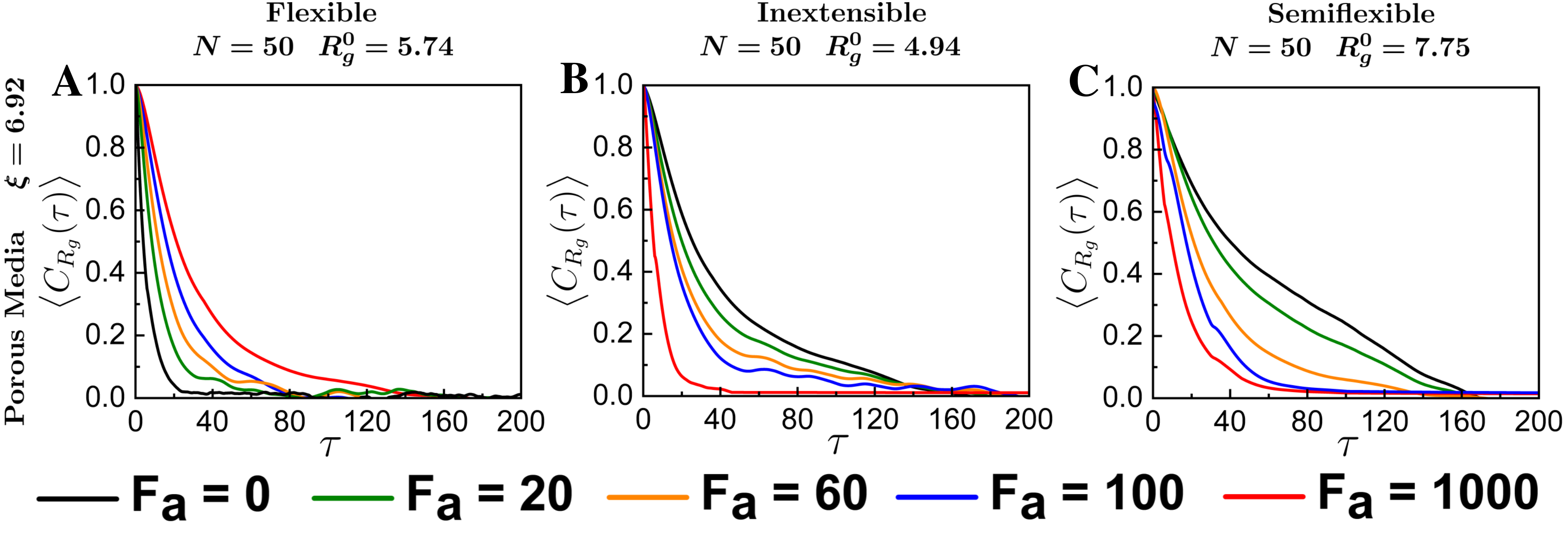} 
\caption{$\left < C_{\text{R}_g}(\tau)\right>$ $vs$ $\tau$ for flexible (A), inextensible (B), and semiflexible (C) ring polymers ($N = 50$) subjected to different activity in the porous medium with $\xi = 6.92$.}\label{fig:Rg_correlation}
\end{figure*}

\noindent To elucidate the sub-diffusive dynamics, we consider the effect of activity on the conformations of ring polymers. In this regard, we compute the probability distribution of radius of gyration, $\textrm{P}(\text{R}_g)$ as a function of $\text{F}_a$ and $\xi$. For a polymer, the radius of gyration is defined as: 
\begin{equation}
R_{g} = \sqrt{ \frac{1}{\text{N}}  \sum_{i=1}^{N} m \Big (r_i - r_{\text{com}} \Big)^2}
\label{eq:gyration}
\end{equation}
where N is the total number of monomers of the ring polymer, and $r_{\text{com}}$ is the center of mass position. $\text{R}_g$ is calculated at each time-step of every simulation after the system reaches the steady-state. Then a single trajectory is created by stitching different individual trajectories. This single trajectory is binned to construct a histogram from which the ensemble averaged probability distribution of $\text{R}_g$ is computed. In an unconfined environment, increasing activity leads to constant swelling of the flexible ring polymer, and hence the peaks of $\textrm{P}(\text{R}_g)$ shift to larger $\text{R}_g$ with $\text{F}_a$ (Fig.~\ref{fig:conformation}A). The reason for the swelling is an increase in the stretching of the chain due to activity by effectively pulling it from different directions. For inextensible ring polymers, there is no shift in the peak values of $\textrm{P}(\text{R}_g)$ because the bond fluctuations of the chain are restricted by the very high value of spring constant in the unconfined space (Fig.~\ref{fig:conformation}B). For a semiflexible ring polymer also, we find that the peaks of $\textrm{P}(\text{R}_g)$ are almost independent of $\text{F}_a$ for the wide range of activities in unconfined space (Fig.~\ref{fig:conformation}C and Fig.~S4A). This implies, for the very high value of bending potential, the bending rigidity dominates over the activity. In contrast to the unconfined space, the flexible ring polymer continuously swells with increasing $\text{F}_a$ even inside the porous medium (Fig.~\ref{fig:conformation}D), but $\left < \text{R}_g\right >$ is smaller as compared to the unconfined space (Fig.~S5). Because, in the unconfined space, the ring can freely change its conformations, while inside the tight pore, the conformational changes are restricted by the obstacles in the media due to limited space available within the obstacles forming the pore confinements. Interestingly, for the inextensible ring polymers, not only does the swelling ceases, but also an opposite behavior is observed in the porous medium. The peaks of the distributions shift towards smaller values of $\text{R}_g$ with an increase in $\text{F}_a$, implying the activity-induced shrinking of the ring in the porous medium (Fig.~\ref{fig:conformation}E and Fig.~S4B). However, a recent study showed that a single active linear polymer undergoes a coil-to-globule like transition in free space~\cite{bianco2018globulelike}. It is important to note that, in their model, active force is applied along the backbone of the chain which pushes the filament along the bond directions, leads to shrinking. In our system, the mechanism of activity-induced shrinking is quite different. In the pore confinement, the ring polymer with higher activity more frequently collides with the obstacles with a larger effective force. This subsequently generates more fluctuations along the inward transverse direction of the contour responsible for the shrinking of ring polymer. The semiflexible ring polymer also exhibits similar features of $\textrm{P}(\text{R}_g)$ as inextensible ring polymers (Fig.~\ref{fig:conformation}F). The activity-induced inward transverse fluctuations attempt to crumple the ring-like structure in the case of inextensible and semiflexible ring polymers. However, this is opposed by the high bending rigidity of the ring polymers when the bending potential is very high. Thus, bending rigidity reduces the extent of shrinking of the semiflexible ring polymer compared to the inextensible ring polymer (Fig.~S5B). We calculate the time autocorrelation functions of $\text{R}_g$, $C_{\text{R}_g}(\tau) = \frac {\left\langle\overline{\text{R}_g^2(t).\text{R}_g^2(t+\tau)}\right\rangle - \left\langle\overline{\text{R}_g^2(t)}\right\rangle^2}{\left\langle\overline{\text{R}_g^4(t)}\right\rangle - \left\langle\overline{\text{R}_g^2(t)}\right\rangle^2}$ to investigate the conformational relaxation of the ring polymer in porous media~\cite{dimitrov2008universal}. For flexible ring, $C_{\text{R}_g}(\tau)$ increases with activity whereas it decreases for the inextensible and semiflexible rings (Fig.~\ref{fig:Rg_correlation}). Hence, the trends in $C_{\text{R}_g}(\tau)$ support the swelling of the flexible ring and shrinking of the inextensible and semiflexible rings in the porous media.  We also plot $1-\textrm{CDF}(\text{R}_g)$ corresponding to each distributions of $\textrm{P}(\text{R}_g)$, where $\textrm{CDF}(\text{R}_g)=\frac{\Sigma_{0}^{\text{R}_g} \textrm{P}(\text{R}_g) \text{R}_g}{\Sigma_{0}^{\infty} \textrm{P}(\text{R}_g) \text{R}_g}$ is the cumulative distribution function of $\text{R}_g$ (Fig.~S6). $1-\textrm{CDF}(\text{R}_g)$ qualitatively demonstrates the similar behavior as observed in $\textrm{P}(\text{R}_g)$. 

\subsection{Nature of the ring determines the migration in the porous media} 

\begin{figure*}
	\centering
		\includegraphics[width=0.95\linewidth]{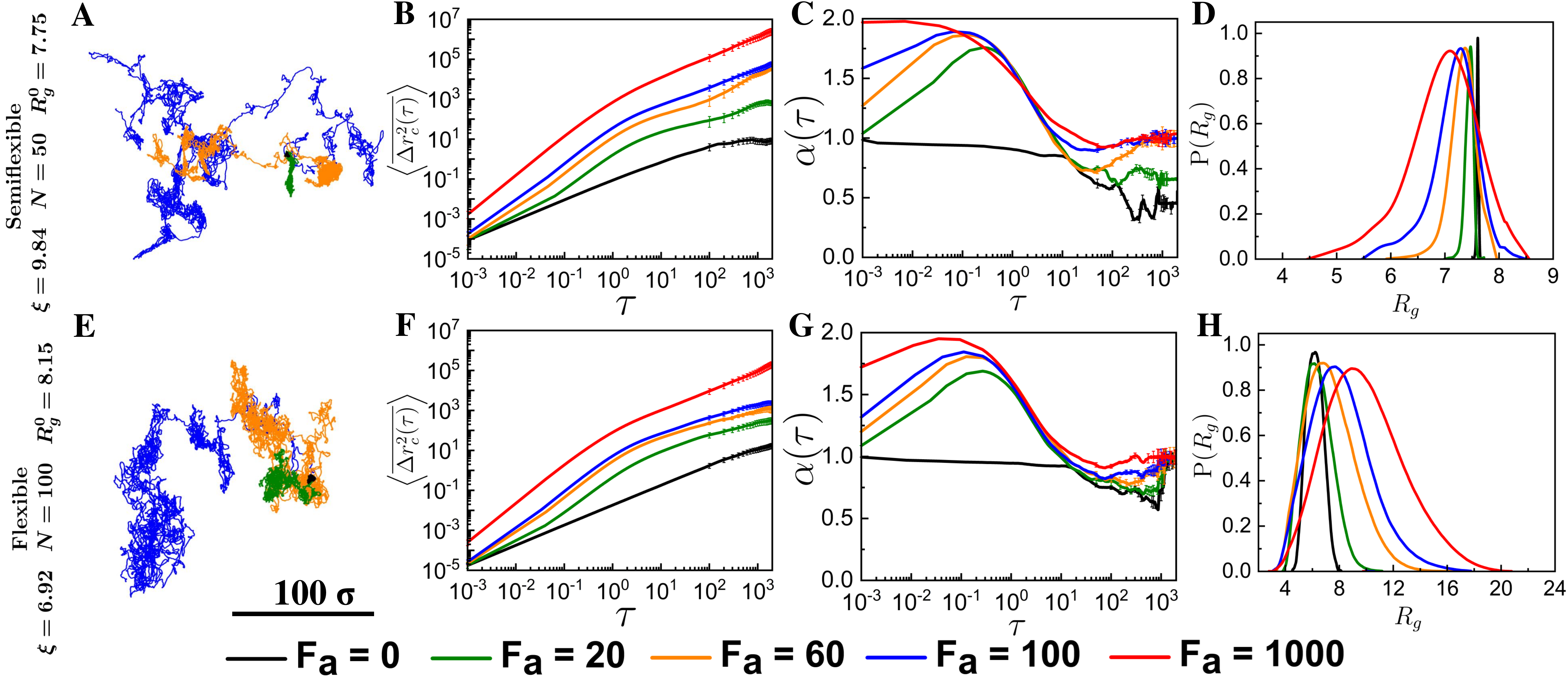} 
\caption{The COM trajectory, log-log plot of $\left\langle{\overline{\Delta r_{c}^{2}(\tau)}}\right\rangle$ $vs$ $\tau$, log-linear plot of $\alpha(\tau)$ $vs$ $\tau$, and $\textrm{P}(\text{R}_g)$ $vs$ $\text{R}_g$ of the ring subjected to different $\text{F}_a$ for flexible (N = 100) (A, B, C, and D) and semiflexible (N = 50) (E, F, G, and H) ring polymers in the porous medium with $\xi = $ 6.92 and 9.84 respectively. For each time interval $\tau$, the error bars corresponding to standard deviation are plotted in $\left\langle{\overline{\Delta r_{c}^{2}(\tau)}}\right\rangle$ (B and F) and $\alpha(\tau)$ (C and G), even if too small to be visible.}\label{fig:compare_N}
\end{figure*}

\noindent Next, we compare the dynamical and conformational properties of active rings in a different fashion. We consider two different cases: a flexible ring of size comparable to the pore size and a semiflexible ring of size smaller than the pore size. The trapping of the semiflexible ring is observed even when the ring is smaller than the pore size (Fig.~\ref{fig:compare_N}(A-C)). However, in the case of the flexible polymer, the ring moves from one pore to another smoothly even if the ring size is comparable to the pore size (Fig.~\ref{fig:compare_N}(E-G)). Hence the trapping of the semiflexible ring and the smooth migration of the flexible ring polymer in porous media are not solely determined by the pore confinement. It depends on the nature of the rings.  The conformational properties are qualitatively similar to Fig.~\ref{fig:conformation}(D and F) even for a flexible ring of size comparable to the pore size and a semiflexible ring of size smaller than the pore size (Fig.~\ref{fig:compare_N}(D and H)).

\section{Discussion}    

\noindent Our simulations characterize how the motion of an active ring polymer in porous media is regulated by both pore-scale confinement and the individual bead's activity. The dynamics of the COM of a ring polymer is enhanced by orders of magnitude and exhibits an intermediate super-diffusive behavior in the presence of activity as compared to the passive ring. Moreover, the semiflexible ring polymer displays a transition from trapping at small activities to escaping at higher activities due to its stretched conformations. According to analysis, the ring polymer shows distinct conformational changes at low and high spring constants in porous environments. Crowders of the porous media induce an unavoidable collapse of the passive ring polymer. However, the activity exerts a force on the bonds attempting to stretch the ring polymers. Hence, upon increasing the activity, the flexible ring polymer swells, while the inextensible ring polymer collapses. The conformational restrictions by higher spring constant increase the inward transverse fluctuations leading to shrinkage of the ring polymer with increasing the activity. Semiflexible active ring polymer also suffers an activity-induced collapse but with a larger average size due to its extended confirmations as compared to the inextensible active ring polymer. \\ 

\noindent The physics underlying the phenomena we report here relies on the motion of a ring polymer driven by the active fluctuations in porous environment. In our model, superdiffusive behavior arises at an intermediate time due to the activity driven breakdown of the fluctuation-dissipation theorem. However, superdiffusive behavior can also occur in a collection of passive elastic ring polymers due to the interplay between the particle deformations and dynamic heterogeneity ~\cite{gnan2019microscopic} . Hence, further studies of collective migration of active ring polymers in complex environments with the inclusion of softness in the monomers ~\cite{gnan2019microscopic} are anticipated in the future.  

\section*{Acknowledgment}

\noindent  L.T. thanks UGC for a fellowship. S.C. thanks DST Inspire for a fellowship. R.C. acknowledges SERB for funding (Project No. MTR/2020/000230 under MATRICS scheme). T.B. acknowledges NCBS-TIFR for research funding. We acknowledge the SpaceTime-2 supercomputing facility at IIT Bombay for the computing time.  

\section*{Data Availability}

\noindent The codes and data used for this paper are available from the authors upon request.

\clearpage

\appendix

\onecolumngrid \section*{\LARGE{Supplementary Material}}

\renewcommand{\arraystretch}{2.0}  
\begin{table*}[b]
\Large
\caption{Model Parameters}
\centering
\begin{tabular}{|c|l|}
\hline
\text{Parameter} &  \multicolumn{1}{|c|}{\text{Value}}\\ \hline
$\sigma$ & 1 \\
$\sigma_P$ & 1 -- 10  \\
$\frac{m}{\gamma}$ & $10^{-4}$ \\
$k_BT$ & 1 \\
$k$ & 30, 1000  \\
$\kappa$ & 1000  \\
$\Delta t$ & $10^{-5}$  \\
$\text{F}_a$ & 0, 20, 60, 100, 1000 \\
$\xi$ & 3.15, 6.92, 12.94   \\
\hline
\end{tabular}
 \label{Tab:para}
\end{table*}

\clearpage
\renewcommand{\thefigure}{S\arabic{figure}}
\setcounter{figure}{0}
\begin{figure*}
	\centering
		\includegraphics[width=0.62\linewidth]{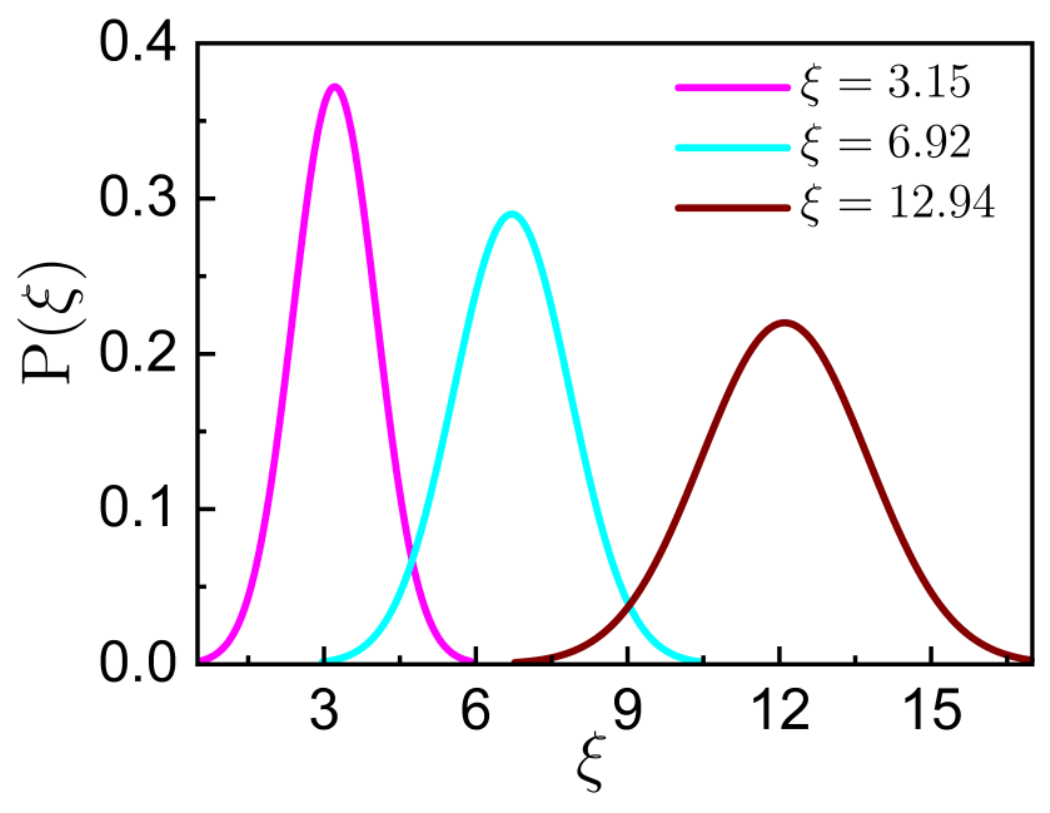} 
\caption{The distribution of pore spaces, $\text{P}(\xi)$ for different porous media.}\label{fig:media_distxi}
\end{figure*}
\begin{figure*}
\centering
   \includegraphics[width=0.99\linewidth]{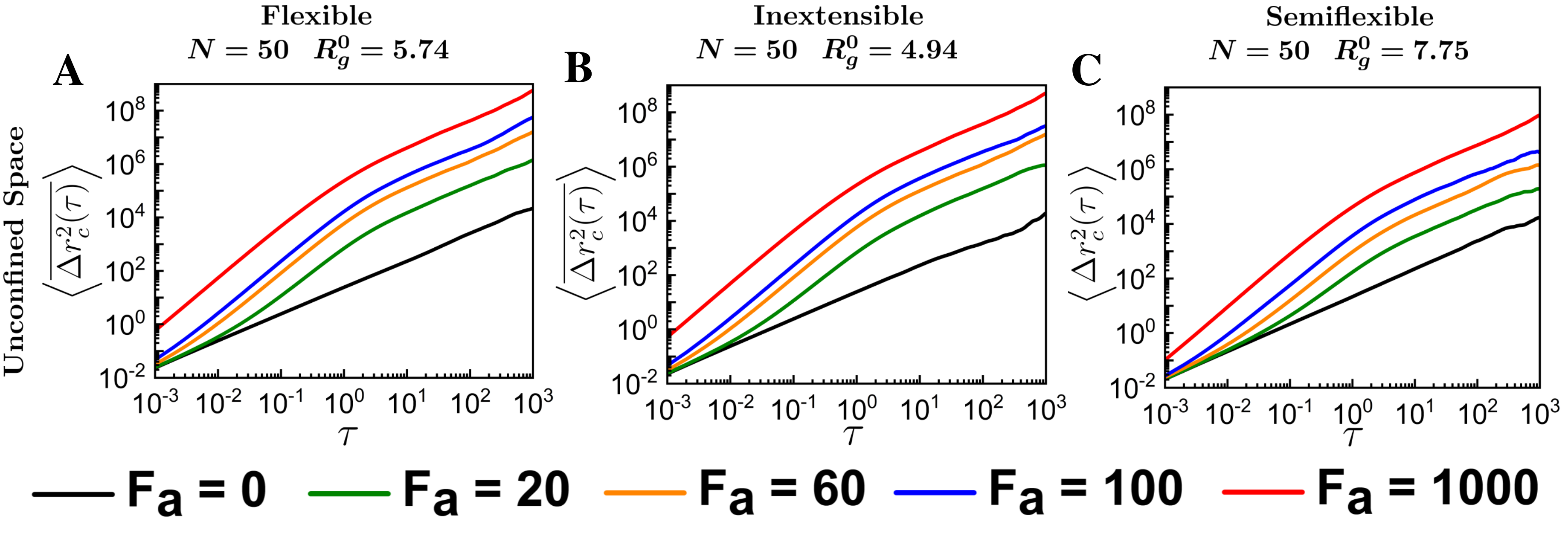} \\
\caption{Log-log plot of $\left\langle{\overline{\Delta r_c^{2}(\tau)}}\right\rangle$ $vs$ $\tau$ for the (A) flexible (B) inextensible and (C) semiflexible ring polymers in unconfined space.}\label{fig:msd_free}
\end{figure*}
\begin{figure*}
	\centering
		\includegraphics[width=0.99\linewidth]{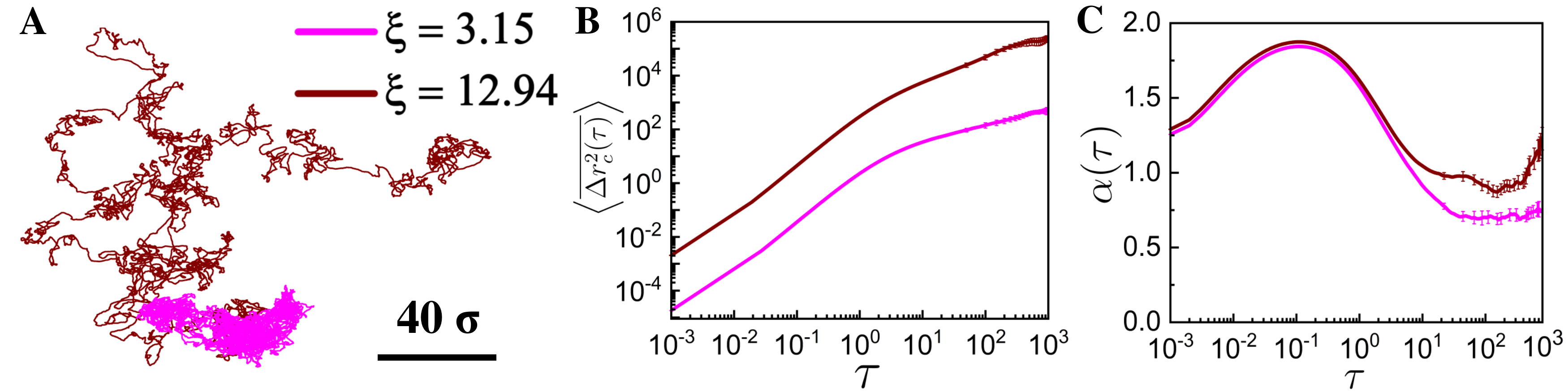} 
\caption{(A) The COM trajectory (B) log-log plot of $\left\langle{\overline{\Delta r_{c}^{2}(\tau)}}\right\rangle$ $vs$ $\tau$ and (C) log-linear plot of $\alpha(\tau)$ $vs$ $\tau$ of the flexible active ($F_a = 60$) ring polymer in different porous media. For each time interval $\tau$, the error bars corresponding to standard deviation are plotted, even if too small to be visible.}\label{fig:media_dynamics}
\end{figure*}
\begin{figure*}
\centering
   \includegraphics[width=0.85\linewidth]{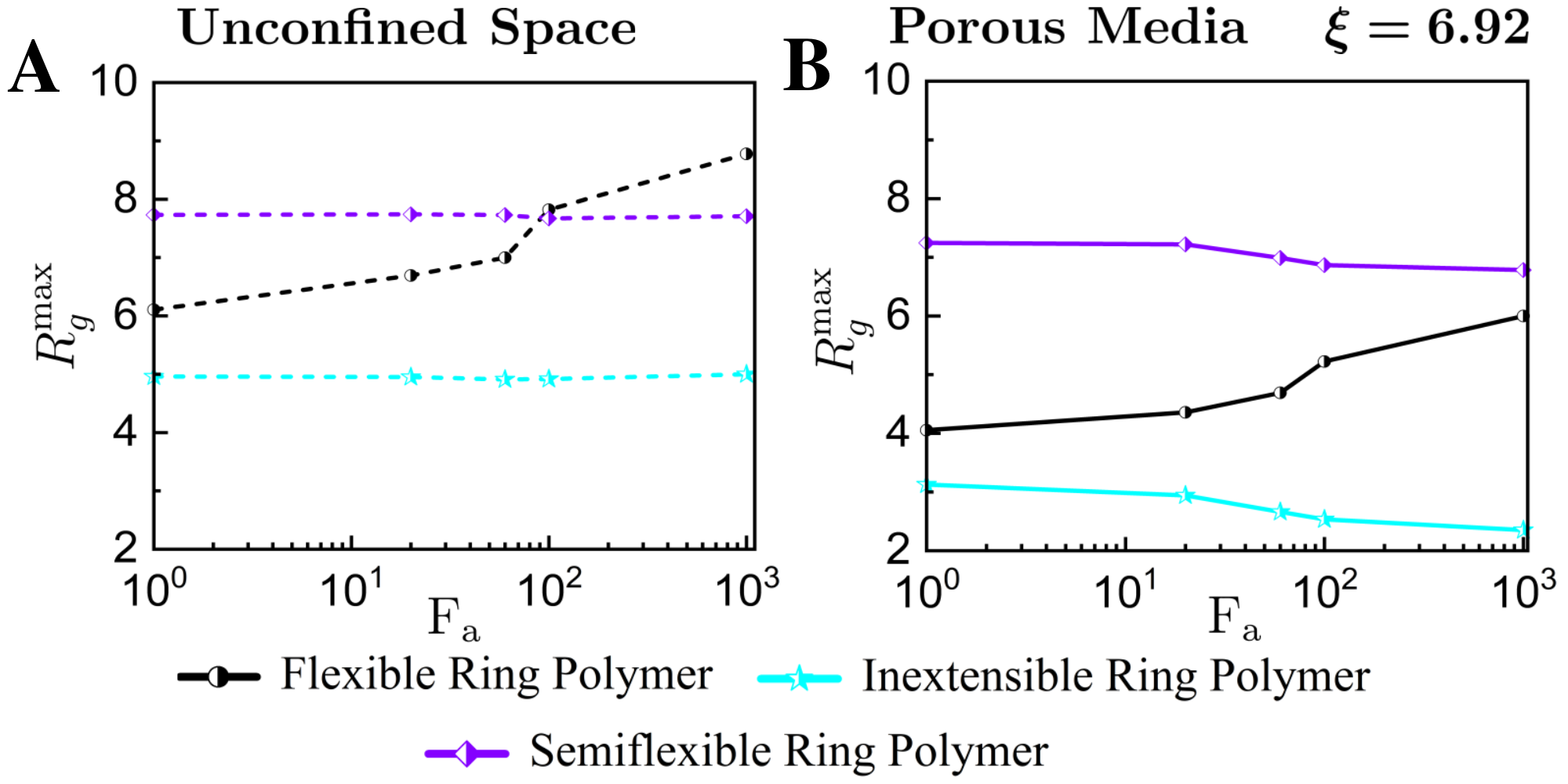} \\
\caption{Log-linear plot of the peak values of $P(R_g)$, $R_g^{\text{max}}$ $vs$ $\text{F}_a$ of the flexible, inextensible, and semiflexible ring polymers in (A) unconfined space (dashed lines) and (B) in porous medium (solid lines) with $\xi = 6.92$.}\label{fig:Rg_Media}
\end{figure*}
\begin{figure*}
\centering
   \includegraphics[width=0.85\linewidth]{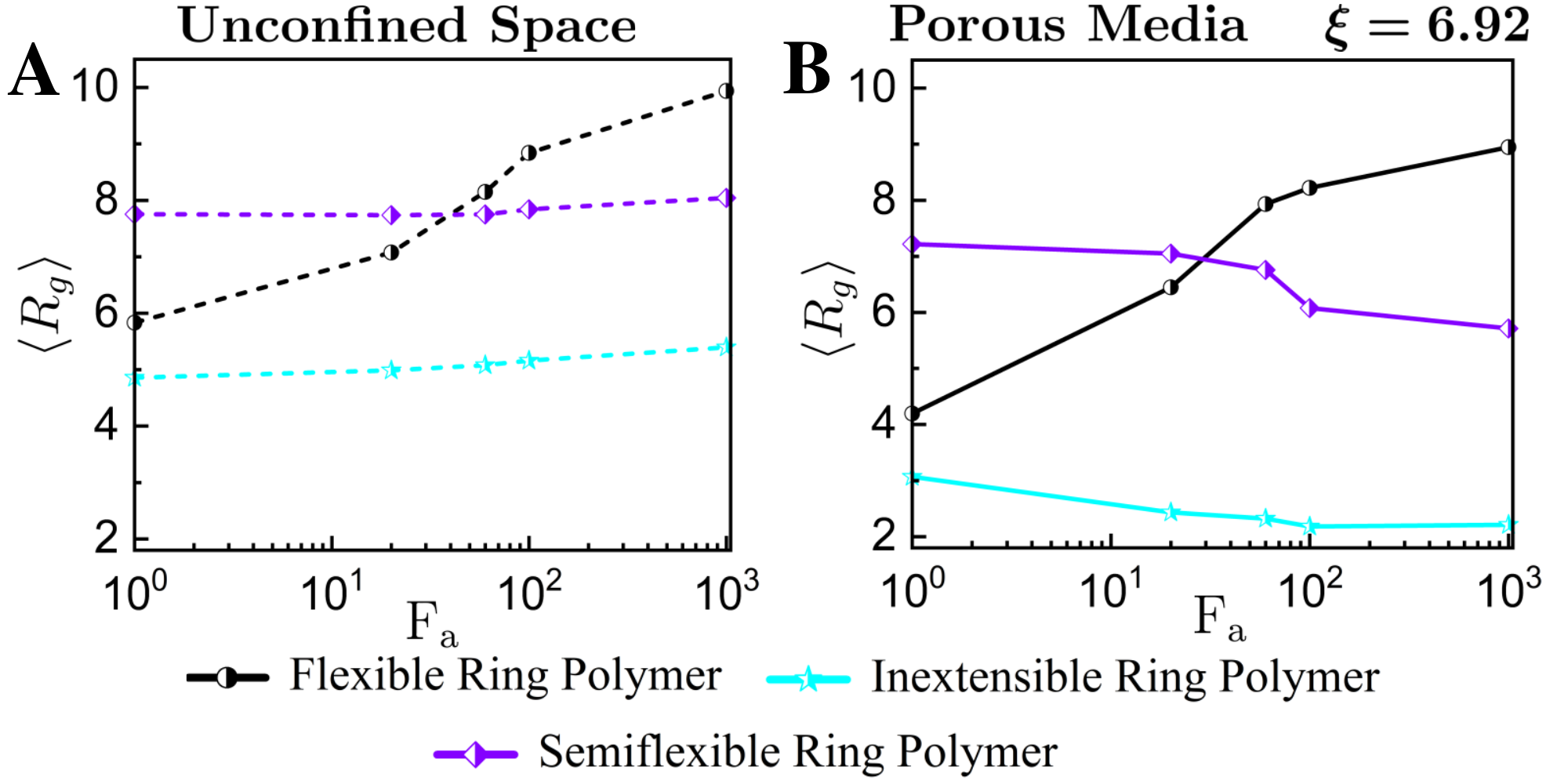} \\
\caption{Log-linear plot of $\left < R_g \right >$ $vs$ $\text{F}_a$ of the flexible, inextensible, and semiflexible ring polymers in  (A) unconfined space (dashed lines) and (B) in porous medium (solid lines) with $\xi = 6.92$.}\label{fig:Rg_Media}
\end{figure*}
\begin{figure*}
\centering
   \includegraphics[width=0.99\linewidth]{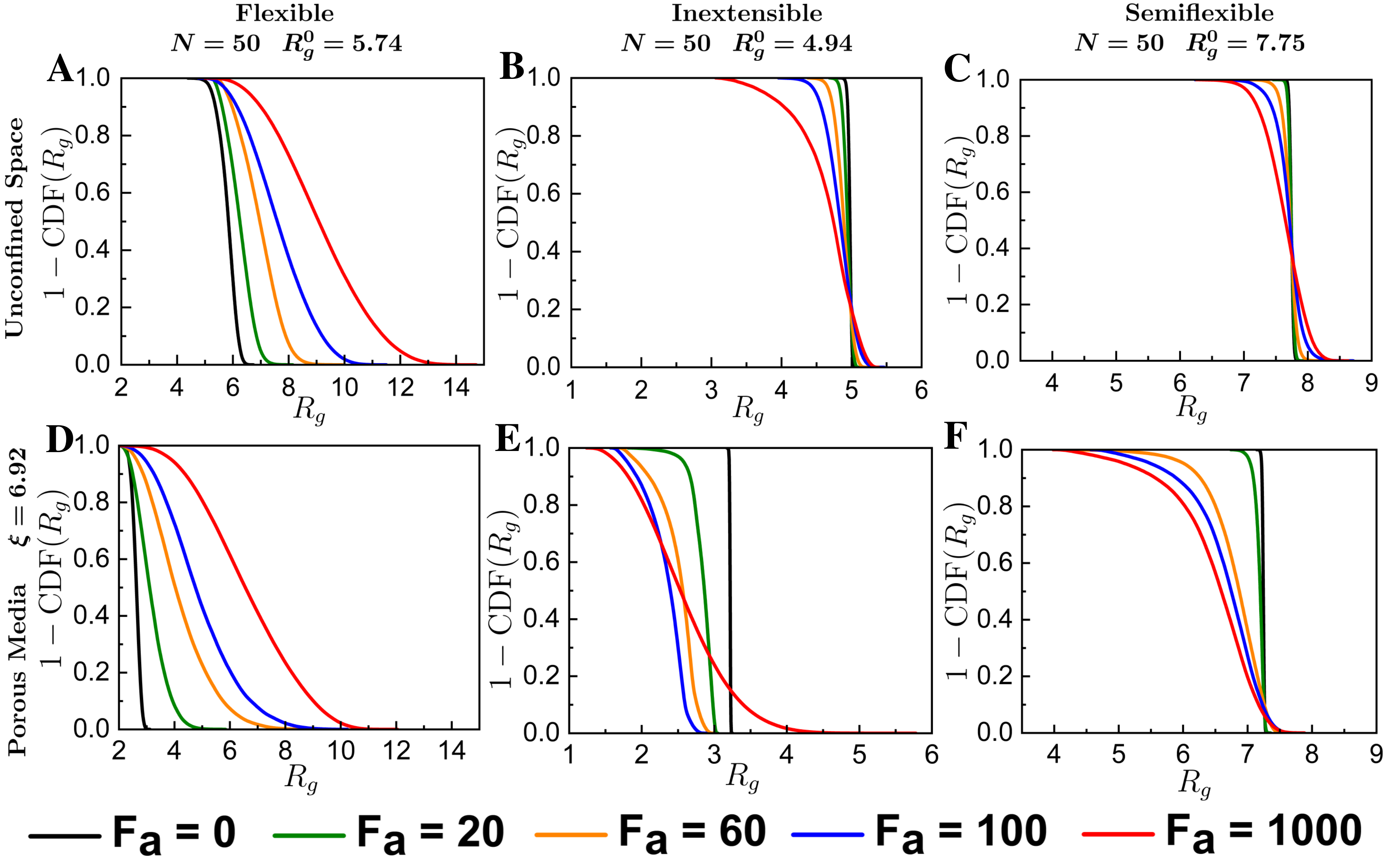} \\
\caption{$1-\textrm{CDF}(R_g)$ $vs$ $R_g$ for flexible (A and D), inextensible (B and E), and semiflexible (C and F) ring polymers ($N = 50$) subjected to different activity in the unconfined space and porous medium with $\xi = 6.92$ respectively. $R_g^0$ is the average $R_g$ of the corresponding passive ring polymer in the unconfined space.}\label{fig:Rg_CDF}
\end{figure*}

\clearpage
\twocolumngrid \subsection{Movie Description}

\noindent \textbf{Movie S1:} The motion of flexible active ($F_a = 60$) ring polymer in random porous media ($\xi = 6.92$). This movie corresponds to the trajectory shown in Fig.~1B. \\ 

\noindent \textbf{Movie S2:} The motion of flexible active ($F_a = 60$) ring polymer in random porous media ($\xi = 6.92$). The ring polymer undergoes a series of conformational changes while navigating through the different pores in the media. \\   

\noindent \textbf{Movie S3:} The motion of inextensible active ($F_a = 60$) ring polymer in random porous media ($\xi = 6.92$). The average size of the inextensible ring polymer is smaller as compared to the flexible (Movie S2) and semiflexible (Movie S4) ring polymers. The inextensible ring polymer effectively squeezes through the porous space by activity-induced shape deformations mainly caused by shrinkage of the ring. \\ 

\noindent \textbf{Movie S4:} The motion of semiflexible active ($F_a = 60$) ring polymer in random porous media ($\xi = 6.92$). The semiflexible ring polymer is bigger compared to the flexible (Movie S2) and inextensible (Movie S3) ring polymers. The semiflexible active ring polymer gets trapped inside smaller pores while moving through the random porous media and escapes from the traps by activity-induced shrinking. The trapping of semiflexible ring polymer is caused by its high bending rigidity, which restricts the conformational fluctuations of the ring polymer. The trapping is not observed for flexible and inextensible ring polymers (Movie S2 and Movie S3). \\

\end{document}